\documentclass{article}

\usepackage{graphicx}				
\usepackage{amsmath}				
\usepackage{amsfonts}
\usepackage{url}
\usepackage[colorlinks=true,linkcolor=blue,anchorcolor=black,citecolor=blue,filecolor=black,menucolor=black,urlcolor=blue,breaklinks=true,pdfhighlight=/P,pdfmenubar=true,pdftoolbar=false,pdfpagelabels=false,pdfstartpage=1,pdfstartview=FitV,pdftitle={On Oscillations in the Social Force Model},pdfsubject={Pedestrian, Crowd, Mass, Simulation, Dynamics, Viswalk, Vissim},pdfauthor={Kretz},pdfcreator={tobias kretz},pdfproducer={t kretz},pdfkeywords={Pedestrian, Crowd, Mass, Simulation, Dynamics, Viswalk}]{hyperref}	
\usepackage[numbers,sort&compress]{natbib}	
\usepackage{hypernat}				
\usepackage[american]{babel}		
\usepackage{simplemargins}
\setallmargins{2cm}
\usepackage[american]{babel}		

\begin{document}

\title{On Oscillations in the Social Force Model}
\author{Tobias Kretz\\
PTV Group, D-76131 Karlsruhe, Germany \\ \texttt{Tobias.Kretz@ptvgroup.com}}

\maketitle

\abstract{The Social Force Model is one of the most prominent models of pedestrian dynamics. As such naturally much discussion and criticism has spawned around it, some of which concerns the existence of oscillations in the movement of pedestrians. This contribution is investigating under which circumstances, parameter choices, and model variants oscillations do occur and how this can be prevented. It is shown that oscillations can be excluded if the model parameters fulfill certain relations. The fact that with some parameter choices oscillations occur and with some not is exploited to verify a specific computer implementation of the model.}


\section{Introduction}
The Social Force Model of pedestrian dynamics is a model that aims at describing the movement of pedestrians with the predominant purpose of simulating pedestrian movement on computers. The force of pedestrian $\beta$ on pedestrian $\alpha$ typically has the form
\begin{equation}
\vec{f}_{\alpha \beta} = A_{\alpha} w() e^{(-g())} \hat{e}_{\alpha \beta}
\end{equation}
where $g()$ is a function which grows with increasing distance between both pedestrians and can depend on the velocities of one or both pedestrians. The function $w()$ suppresses forces the more pedestrian $\beta$ is located outside the current walking direction of pedestrian $\alpha$. 

The Social Force Model has first been introduced in 1995 \cite{helbing1995social}. This variant later was called ``elliptical specification I''. A second variant (circular specification) has been proposed in 2000 \cite{helbing2000simulating} and a third variant (elliptical specification II) in 2007 \cite{Johansson2007}. The difference between the three variants lies mainly in the way the velocities of two interacting pedestrians are considered in the computation of the force between them. The 1995 variant considers only the velocity of the pedestrian who exerts the force. The 2000 variant does not at all consider velocities (only the distance between pedestrians) and the 2007 variant considers the relative velocity between both pedestrians (the pedestrian who exerts the force and the pedestrian on whom the force acts). For the analytical considerations in this paper mainly the simplest variant from 2000 will be considered. Nevertheless it will also be discussed how results will change qualitatively if the variants as of 1995 or 2007 are applied.

Under ``oscillations'' in this paper unrealistic artifacts in the trajectory of a pedestrian approaching another pedestrian, his destination or a wall is understood. The occurrence of oscillations in the sense of this contribution has been discussed in a number of contributions \cite{steffen2008repulsive,chraibi2010generalized,chraibi2013force,koster2013avoiding} and it is often claimed that oscillations cannot be avoided in the Social Force Model, but that they just can be made small. In this paper it will be shown that this is not correct and exact conditions for the value of the model parameter such that oscillations occur will be derived.

In the remainder of the paper first a single pedestrian approaching a destination is investigated, then a pedestrian approaching another pedestrian who is standing still and finally two pedestrians approaching each other. In each case the model is reduced to one dimension and the noise term is set to zero. In the first section on approaching a destination the problem will be shown to be most severe as with certain conditions oscillations cannot be prevented and continue infinitely long. At the same time -- as will be argued -- for this case it is not very relevant, as there are simple, pragmatic solutions. The second and third case yield restrictions to the choice of parameters which can produce realistic, oscillation-free behavior.

\section{A pedestrian approaching a destination}
In this section we are interested in and discuss the equations of motion and their solution of a single pedestrian approaching a destination coordinate (i.e. a point) where he is required to come to a standstill. We assume that in the beginning the pedestrian is walking with his desired speed $v_0$ straight towards the destination coordinate, so there is no tangential component of the walking velocity. Then we can describe the pedestrian as walking from positive $x$ coordinate into negative $x$ direction towards the destination which is at $x=0$. Since the 1995, 2000, and 2007 variants of the Social Force Model only differ in the force between pedestrians and not the driving force term all results of this section hold for all three variants.

We assume for now, that the desired velocity is always some externally given $v_0$ and is always pointing from the pedestrians current position towards $x=0$. This assumption is the simplest one and it can be questioned -- as we will do below. With it it is obvious that there will be oscillations around $x=0$. Our intention here is to investigate the quantitative details of these oscillations.

In general the equation of motion for this pedestrian reads
\begin{equation}
\ddot{x}(t)=\frac{-sign(x(t)) v_0 -\dot{x}(t)}{\tau}
\end{equation}
where $\tau$ is an external parameter which typically has values between 0.1 and 1.0 seconds.

We require the pedestrian not only to reach $x=0$, but also to stand still there as arrival condition. Because the pedestrian has a speed larger 0 (or, considering walking direction: smaller 0) he will walk over the destination and be on the left (negative) side of $x$ coordinates. There the desired velocity points into the direction of positive x coordinates. So we have for the time following the moment when the pedestrian is at $x=0$:
\begin{equation}
\ddot{x}(t)=\frac{v_0 - \dot{x}(t)}{\tau}
\end{equation}

This is solved by 
\begin{eqnarray}
\dot{x}(t) &=& v_0 - a e^{-\frac{t}{\tau}} \label{eq:dotx}\\
x(t)       &=& b + v_0 t + a \tau e^{-\frac{t}{\tau}} \label{eq:x}
\end{eqnarray}
where $a$ and $b$ are integration constants which need to be determined by initial conditions.

We choose $t=0$ at the moment when the pedestrian is at $x=0$. Then $\dot{x}(t=0)=-v_0$. However, for later usage we want to set here more general $\dot{x}(t=0)=-u$ and remember that for our particular case $u=v_0$. With the two conditions $x(0)=0$ and $\dot{x}(0)=-u$ we can determine the values of the integration constants:
\begin{eqnarray}
         a&=&v_0+u\\
         b&=&-(v_0+u)\tau
\end{eqnarray}

So we have
\begin{eqnarray}
\dot{x}(t)&=&v_0-(v_0+u) e^{-\frac{t}{\tau}}\\
x(t)&=&v_0t-(v_0+u)\tau \left(1- e^{-\frac{t}{\tau}} \right)
\end{eqnarray}

Now we can compute the time $t_{turn}$ when the pedestrian stops (and turns around) $\dot{x}(t_0)=0$ and the position $x(t_0)$ at which this happens:
\begin{eqnarray}
              t_{turn}&=&\tau \ln\frac{v_0+u}{v_0}\\
x(t_{turn})&=&\tau v_0 \left(\ln\left(1+\frac{u}{v_0}\right)-\frac{u}{v_0}\right) \label{eq:xt0}
\end{eqnarray}

In the initial case, when $u=v_0$ this simplifies to $t_{turn}=\tau \ln(2)$ and $x(t_{turn})=\tau v_0 (\ln(2) - 1)$.

This is only half the way to go. The actual question is how fast a pedestrian returns to $x=0$ when he has passed it before with speed $u$ and how long such a loop takes.

Now we choose $t=0$ for the moment when the pedestrian is standing still at $x(t_{turn})$. The time $t_{returned}$ at which the pedestrian returns to $x=0$ therefore has to be understood as time after $t_{turn}$ and not as absolute point in time.

We begin again by determining the value of the integration constants. From equation (\ref{eq:dotx}) and $\dot{x}(0)=0$ follows $a=v_0$. With equations (\ref{eq:x}) and (\ref{eq:xt0}) we have
\begin{eqnarray}
x(0)&=&b+v_0\tau = v_0\tau \left(\ln\left(1+\frac{u}{v_0}\right)-\frac{u}{v_0}\right)\\
b&=&v_0\tau \left(\ln\left(1+\frac{u}{v_0}\right)-\left(1+\frac{u}{v_0}\right)\right)\\
x(t)&=&v_0\tau\left(\ln\left(1+\frac{u}{v_0}\right)-\left(1+\frac{u}{v_0}\right)+\frac{t}{\tau} + e^{-\frac{t}{\tau}}\right)
\end{eqnarray}

Because $x(t_{returned})=0$ we have the following equation for $t_{returned}$:
\begin{eqnarray}
0&=&v_0\tau\left(\ln\left(1+\frac{u}{v_0}\right)-\left(1+\frac{u}{v_0}\right)+\frac{t_{returned}}{\tau} + e^{-\frac{t_{returned}}{\tau}}\right)\\
0&=&\phi_{returned}-\alpha+\ln(\alpha) + e^{-\phi_{returned}}
\end{eqnarray}
with the substitutions $\phi_{returned}=t_{returned}/\tau$ and $\alpha=1+u/v_0$. We do another substitution $\varphi=\phi_{returned}-\alpha+\ln(\alpha)$, transforming the last equation to

\begin{eqnarray}
0&=&\varphi+e^{-(\varphi+\alpha-\ln(\alpha))}\\
-\alpha e^{-\alpha}&=&\varphi e^{\varphi}
\end{eqnarray}

Obviously one solution is $\varphi = -\alpha$, but the general solution is
\begin{equation}
\varphi=W\left(-\alpha e^{-\alpha}\right)
\end{equation}
where $W()$ is the Lambert $W$ function \cite{corless1996,corless1997,chapeau2002}, which by definition is the inverse relation of $f(y)=ye^y$. 

Resubstituting $\varphi$ and $\phi_{returned}$ (for reasons of convenience we do not resubstitute $\alpha$) we have:
\begin{equation}
\frac{t_{returned}}{\tau}=W\left(-\alpha e^{-\alpha}\right)+\alpha-\ln(\alpha) \label{eq:t1}
\end{equation}

In the interval $y \in [-1/e..0]$ $W(y)$ has two branches, denoted $W_{-1}(y)$ and $W_0(y)$; and with $1 \leq \alpha \leq 2$ for $-\alpha e^{-\alpha}$ we are within this interval where $W(y)$ has two branches. It holds
\begin{equation}
W_{-1}\left(-\alpha e^{-\alpha}\right)=-\alpha \label{eq:W1W0}
\end{equation}
In this case it would be $t_{returned}=-t_{turn}$. Thus the $W_{-1}$ branch gives the backwards in time solution which we are not interested in here (because we already have computed it above). Therefore we have to continue with the $W_0$ branch for which 
\begin{equation}
W_{0}\left(-\alpha e^{-\alpha}\right)\neq -\alpha
\end{equation}
although $W$ is the inverse relation of $f(y)=y e^y$. In the remainder we write $W$ for $W_0$.

In case $u=v_0$, i.e. $\alpha=2$ the numerical value of the solution is $\frac{t_{returned}}{\tau}=W(-2/e^{2})+2-\ln(2)=0.90047$.

Using equation (\ref{eq:t1}) in equation (\ref{eq:dotx}) we get the speed $\dot{x}(t_{returned})$ at the time when the pedestrian returns to $x=0$ in dependence from $u$, which is the speed at the time when the pedestrian last was at $x=0$:
\begin{equation}
\dot{x}(t_{returned})=v_0\left(1+W\left(-\alpha e^{-\alpha}\right)\right)\label{eq:vreturned}
\end{equation}
where we have used the defining equation of the $W$ function $y=W(y)e^{W(y)}$.

From here on the properties have an index that determines the recurrence to the offspring. For example $t_0$ is the (absolute) time when the pedestrian is for the first time at the offspring, $t_1$ denotes the (absolute) time when the pedestrian returns for the first time to the offspring and so on. In the case of properties that do not describe passage of the offspring, but the loop durance and turning position the index enumerates the loops. This means that for these two properties there is no index 0 and that $t_{n}=t_{n-1} + \Delta t_{n}$.

Equation (\ref{eq:vreturned}) means that for the $n+1$th passage of the offspring $\alpha_{n+1}$ depends on $\alpha_n$ like
\begin{equation}
\alpha_{n+1}=2+W\left(-\alpha_n e^{-\alpha_n}\right)\label{eq:alphanp1}
\end{equation}

The time $\Delta t_{n+1}$ it takes for the $n+1$th loop depends on $\alpha_n$ like
\begin{equation}
\frac{\Delta t_{n+1}}{\tau} = \alpha_n + W\left(-\alpha_n e^{-\alpha_n}\right). \label{eq:Dt}
\end{equation}

Rewriting equation (\ref{eq:xt0}) in terms of $\alpha$ and writing $|x_n|$ for the turn around distance of the $n$th loop we have
\begin{equation}
\frac{|x_{n+1}|}{\tau v_0}=\alpha_n-1-\ln(\alpha_n)
\end{equation}

\begin{figure}[htbp]%
\center
\includegraphics[width=0.45\columnwidth]{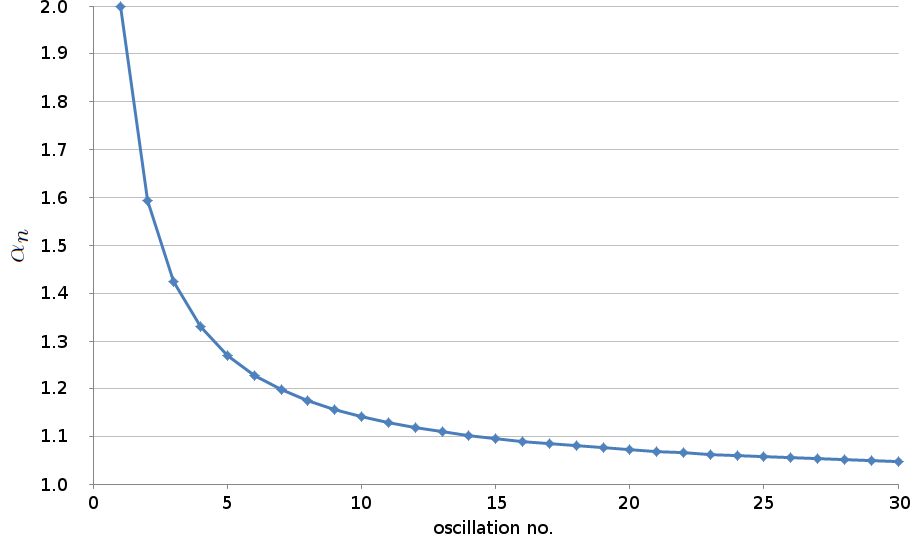} \hspace{12pt}%
\includegraphics[width=0.45\columnwidth]{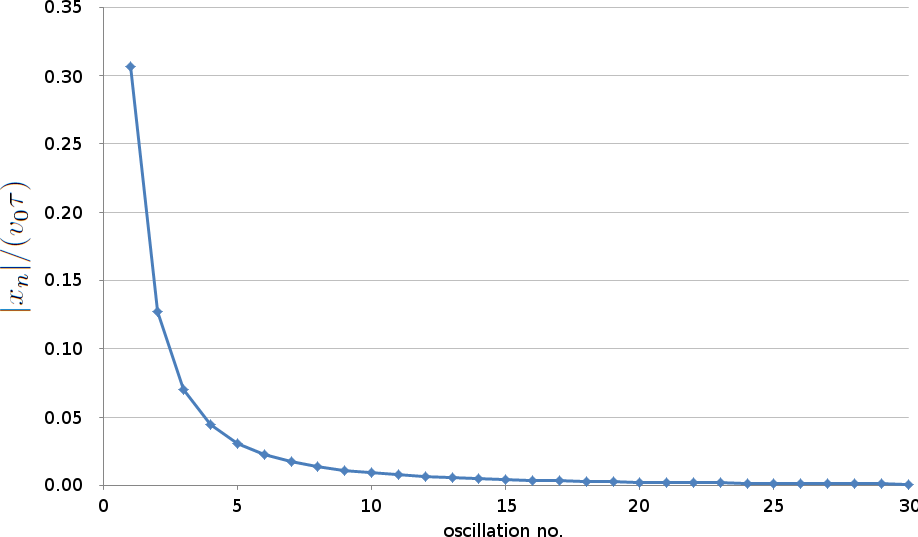}%
\caption{Left: Value of $\alpha_n$ at the beginning of oscillation $n$. Right: Maximum amplitude $|x_n|/(v_0\tau)$ of oscillation $n$.}%
\label{fig:alphan-and-xn}%
\end{figure}

\begin{figure}[htbp]%
\center
\includegraphics[width=0.45\columnwidth]{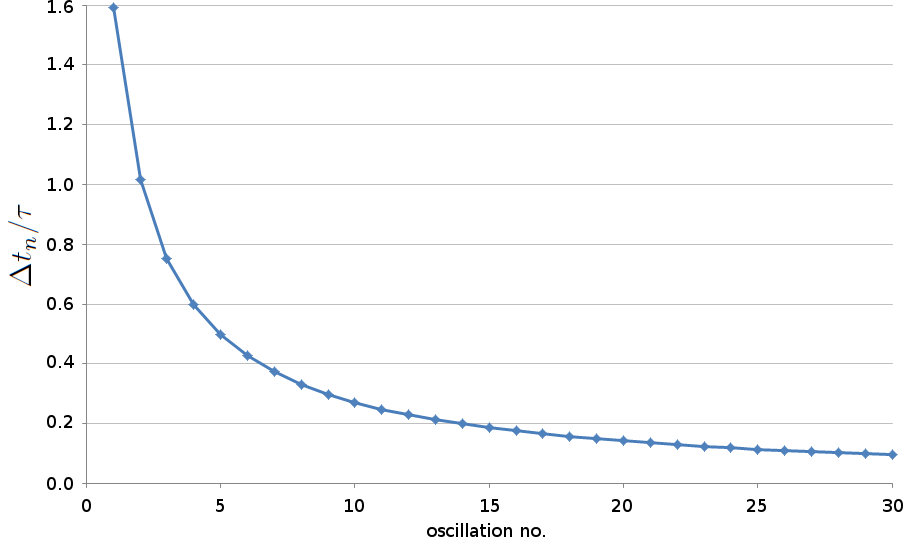} \hspace{12pt}%
\includegraphics[width=0.45\columnwidth]{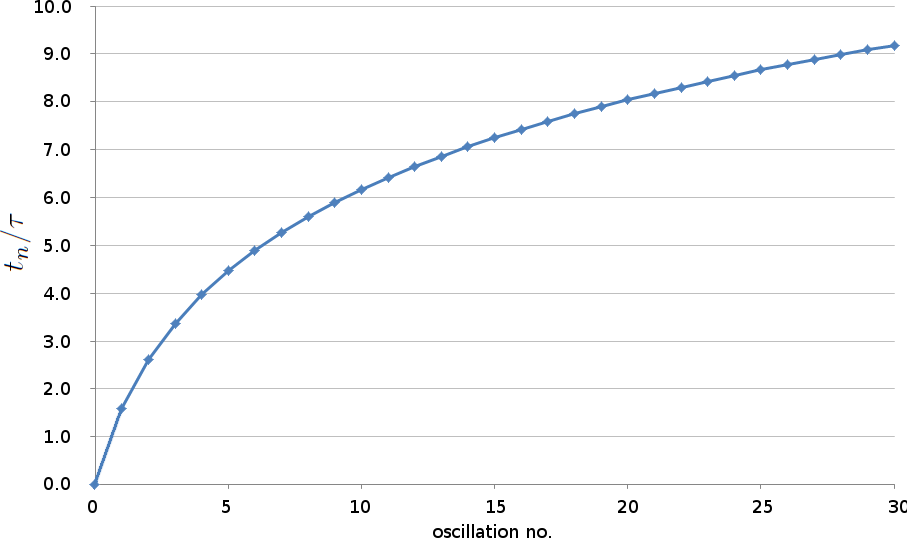}%
\caption{Left: Period $\Delta t_n / \tau$ of oscillation $n$. Right: Total time  $t_n / \tau$ at $n$th passage of offspring (i.e. time after $n$th oscillation; t=0 is when the pedestrian reaches the offspring for the first time and with a speed $v_0$).}%
\label{fig:Dtn-and-tn}%
\end{figure}

\begin{figure}[htbp]%
\center
\includegraphics[width=0.612\columnwidth]{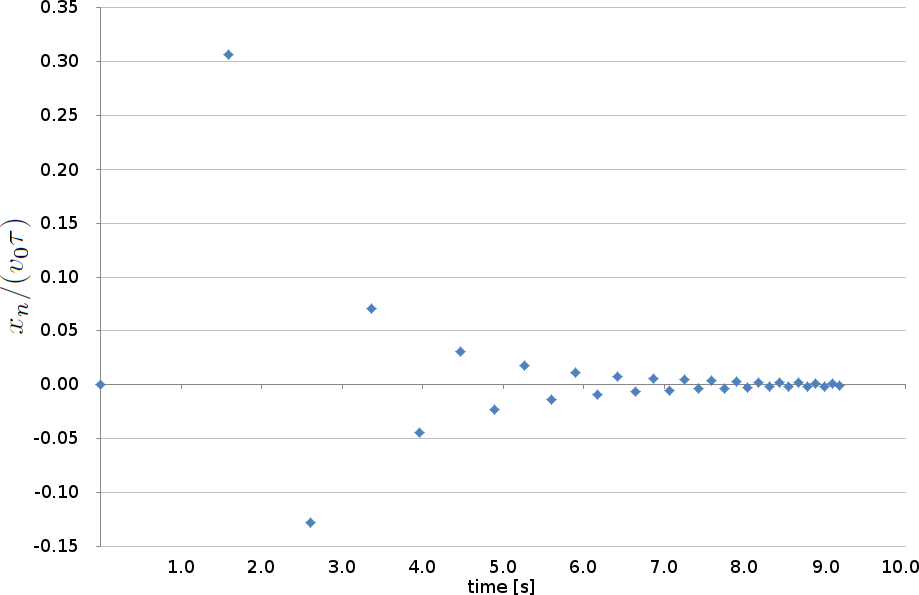}%
\caption{This shows the evolution of the amplitude $x_n/(v_0\tau)$ over time in [s].}%
\label{fig:x-n---t-n}%
\end{figure}

Table \ref{tab:results1} shows the results from the first 30 passages if the process begins at $t=0$ and with speed $v_0$. Figures (\ref{fig:alphan-and-xn}) to (\ref{fig:x-n---t-n}) visualize these data. 

\begin{table}
\center
\begin{tabular}{r|ccccc}
Passage/Loop($n$) &  $\frac{|x_n|}{\tau v_0}$ &  $\frac{\Delta t_n}{\tau}$  &  $\frac{t_n}{\tau}$ & $\frac{u_n}{v_0}$ &  $\alpha_n$\\ \hline
 0 &   -   &    -  & 0.000 & 1.000 & 2.000 \\
 1 & 0.307 & 1.594 & 1.594 & 0.594 & 1.594 \\
 2 & 0.128 & 1.018 & 2.611 & 0.424 & 1.424 \\
 3 & 0.071 & 0.754 & 3.365 & 0.330 & 1.330 \\
 4 & 0.045 & 0.600 & 3.966 & 0.270 & 1.270 \\
 5 & 0.031 & 0.499 & 4.465 & 0.229 & 1.229 \\
 6 & 0.023 & 0.428 & 4.893 & 0.199 & 1.199 \\
 7 & 0.017 & 0.374 & 5.266 & 0.175 & 1.175 \\
 8 & 0.014 & 0.332 & 5.599 & 0.157 & 1.157 \\
 9 & 0.011 & 0.299 & 5.898 & 0.142 & 1.142 \\
14 & 0.005 & 0.199 & 7.062 & 0.096 & 1.096 \\
19 & 0.003 & 0.150 & 7.898 & 0.073 & 1.073 \\
29 & 0.001 & 0.100 & 9.088 & 0.049 & 1.049 \\ \hline
\end{tabular}
\caption{Numerical results for the first 30 passages resp. oscillations. The value for $\alpha_n$ is understood {\em before} an oscillation, while $\Delta t_n / \tau$ is the period of that oscillation, and the total time $t_n/\tau$ {\em after} that same oscillation. All values are dimensionless. The numerical values of the Lambert $W$ function have been computed using Wolfram Alpha  \cite{wolframalpha}.}
\label{tab:results1}
\end{table}

As typical values are $\tau=0.4$ s and $v_0=1.5$ m/s it takes about 7 passages before the amplitude $|x_n|$ gets smaller than 1 cm. This takes about 2.1 seconds. The 7th oscillation has a period of 0.15 seconds. To resolve this a computer implementation would have to have a time step which is at the very maximum half that value 0.075 seconds or 14 simulation steps per second.

Because $\alpha_{n+1}=1$ only if $W(-\alpha_n e^{-\alpha_n})=-1$ which in turn is only the case if $\alpha_{n}=1$ there will be infinitely many oscillations and in \ref{app:convergence} a proof is given that the sum of the $\Delta t_n$ diverges, i.e. the pedestrian will oscillate infinitely long around the destination coordinate.

At first sight this is an unsatisfactory result with regard to the Social Force Model, as such oscillations are unrealistic. However, we have to ask how realistic our assumptions and initial conditions were. This concerns in particular the desired velocity $\vec{v}_0$. We demanded in the beginning that the pedestrian should come to a stop at $x=0$. Nevertheless we have set the {\em desired} velocity all of the time to one particular value $|\vec{v}_0|>0$. This is too simplistic. Real persons plan ahead and adjust their desired speed: if they desire to stop at a certain position they adapt their desired velocity beforehand to just achieve that. So we have to ask how we have to modify $v_0$ dynamically to account for that.

More precisely we ask: at which distance $d_b$ does a pedestrian have to start braking in the Social Force Model, if he sets $v_b$ as desired speed opposing his current speed $u$? And how long does it take before he comes to a stand still?

For this the equation of motion reads
\begin{equation}
\dot{x}(t)=\frac{v_b-\dot{x}(t)}{\tau}
\end{equation}

with the following (initial) conditions:
\begin{eqnarray}
\dot{x}(t=0)&=&-u\\
x(t=0)&=&d\\
\dot{x}(t_0)&=&0\\
x(t_0)&=&0
\end{eqnarray}
where we have defined the moment where the pedestrian starts to brake as $t=0$ and the moment where he stops at $x=0$ as $t_0$; $d$ is the distance before stand still at which braking has to begin which we are looking for.

Solving this results in
\begin{eqnarray}
t_0&=&\tau \ln\left(1+\frac{u}{v_b}\right)\\
d_b&=&\tau v_b \left(\frac{u}{v_b} - \ln \left(1+\frac{u}{v_b}\right)\right)
\end{eqnarray}

This gives finite and positive $t_0$ and $d_b$ for all $v_b>0$. Thus, it is not sufficient to set the desired speed to zero for a pedestrian who wants to stop as this would imply an infinitely large braking distance and an infinitely long time to come to a stand still. If we assume that a pedestrian for braking can at maximum have a desired speed $v_b=v_0$ the minimum time for braking is $t_0=\tau \ln(2)$ and the minimum braking distance is $d_b=\tau v_0 (1 - \ln (2))$ to come to standstill from an initial speed $u=v_0$.

A different and pragmatic solution of the problem is that in real-world planning applications usually there is no requirement set for the speed at which a pedestrian arrives at a destination. It is furthermore usual that as destination not a point is given but that a cross section has to be crossed or the pedestrian has to move into a specific area. If no restriction is given for the speed at arrival the problem disappears.

A third objection is that also a real person cannot comply to the request to stand still with his or her center exactly at some given coordinate. Real people always sway a little bit when they (try to) stand still \cite{winter1995human}. Why then should one require this from a simulated pedestrian? The oscillations around the destination point which were found here in their functional form surely do not match the swaying of real people when they stand ``still'', but the amplitude of the ones of the model quickly fall below those of real people. When the real swaying is not required to be reproduced by models implicitly a required limit to precision is set and any movement of that order of magnitude or below should be acceptable.

\section{A pedestrian approaching a standing pedestrian}
\subsection{Theory}
Now we want to investigate the situation when some pedestrian M is approaching another (pedestrian S) one who is standing at a fixed position -- again in one dimension, for an empirical investigation see for example \cite{gorrini2014experimental}. Since the pedestrian who is exerting the force is not moving, the 1995 variant (elliptical specification I) of the Social Force Model will produce the same result as the 2000 variant (circular specification) which is the one we are investigating, but different from the 2007 variant (elliptical specification II) where the speed of the pedestrian on whom the force acts modifies the strength of the force.

Assume pedestrian S is standing still at $x=0$, facing into positive $x$ direction and pedestrian M is approaching from there, i.e. pedestrian M has positive $x$ coordinate and a speed directed towards negative $x$, i.e. $\dot{x}(t=0)<0$ as well as the desired speed $v_0>0$ which is constant in time. It is assumed that pedestrian M at time t=0 is far away from pedestrian S and walking with his desired walking speed $\dot{x}(t=0)=-v_0$.

Then the equation of motion for pedestrian M in the 1995 and 2000 variants of the Social Force Model is
\begin{equation}
\ddot{x}(t)=\frac{-v_0-\dot{x}(t)}{\tau}+Ae^{-\frac{x(t)-2R}{B}} \label{eq:basic}
\end{equation}
with $R$ as radius of a pedestrian (we assume here for simplicity that all pedestrians have the same radius) and $\tau$, $A$, and $B$ parameters of the Social Force Model. 

Pedestrian M will come to a standstill at distance
\begin{equation}
d_s=B\ln\left(\frac{A\tau}{v_0}\right)+2R \label{eq:ds}
\end{equation}
As pedestrians should not bump one into another $d_s$ must be larger than $2R$. Therefore we have a first condition for the parameters of the Social Force Model to yield realistic results:
\begin{equation}
A\tau>v_0 \label{eq:cond1}
\end{equation}

We assume that oscillations only occur generally if they also occur close to equilibrium. Thus, if there are no oscillations when pedestrian M is already close to $x=d_s$ then there are no oscillations at all. Expanding the exponential function in equation (\ref{eq:basic}) into a Taylor series around $x=d_s$ gives
\begin{equation}
\ddot{\xi}(t)+\frac{\dot{\xi}(t)}{\tau}+\frac{v_0}{B\tau}\xi(t) \approx 0 \label{eq:dho}
\end{equation}
Equation (\ref{eq:dho}) is the equation of the {\em damped harmonic oscillator}, with the known three solutions: under damped (pedestrian M approaches the equilibrium point oscillating around it), critically damped (pedestrian M approaches the equilibrium point as quick as possible but without oscillations) and over damped (pedestrian M approaches the equilibrium point slower than would be possible without having oscillations. Which of the three cases is realized depends on the relation of parameters:
\begin{eqnarray}
4\frac{v_0\tau}{B}&>&1 \leftrightarrow \text{ under damped} \\
4\frac{v_0\tau}{B}&=&1 \leftrightarrow \text{ critically damped} \\
4\frac{v_0\tau}{B}&<&1 \leftrightarrow \text{ over damped}
\end{eqnarray}

Thus in addition to equation (\ref{eq:cond1}) with 
\begin{equation}
4v_0\tau\leq B \label{eq:cond2}
\end{equation}
we have a second condition for the parameters of the 1995 and 2000 variants of the Social Force Model to have a chance to yield realistic results.


\subsection{Implications for parameters found in literature}
In the literature one can find values for some or all of the parameters $A$, $B$, $\tau$, and $v_0$ gained from empirical observations or laboratory experiments. If all four parameters are given, one can use relations (\ref{eq:cond1}) and (\ref{eq:cond2}) to test if realistic results for the scenario discussed here can be expected. If not for all four parameters values are given those two equations allow to set limits. Table \ref{tab:literaturecomparison} shows a compilation of such data and how they relate to or with relations (\ref{eq:cond1}) and (\ref{eq:cond2}). It can be seen that where all four parameters were taken from experiment relation (\ref{eq:cond1}) easily holds while relation (\ref{eq:cond2}) is clearly violated. Whereas if parameters $A$ and $B$ (and $\tau$) have been calibrated or given in a work then relations (\ref{eq:cond1}) and (\ref{eq:cond2}) require pedestrians to walk quite slowly, while if $v_0$ and $\tau$ are calibrated or given it needs relatively large values for $A$ and $B$ that relations (\ref{eq:cond1}) and (\ref{eq:cond2}) hold. This is an indication that with the circular specification (alone) the parameter space that yields realistic results is small.

\begin{table}
\center
\begin{tabular}{c|cccccc}
Source                      & $A$ in [$m/s^2$]                  & $B$ in [$m$]          & $\tau $ in [$s$]         & $v_0$ in [$m/s$]      & eq. (\ref{eq:cond1}) & eq. (\ref{eq:cond2}) \\ \hline
\cite{Johansson2007}        & {\bf 0.42 $\pm$ 0.26}             & {\bf 1.65 $\pm$ 1.01} & $\tau>v_0/A_{max}$       &  $<0.67$              &    &   \\
                            &                                   &                       & $\tau\leq B_{max}/v_0/4$ &                       &    &   \\
\cite{helbing2000simulating}& {\bf 26.67}$^\dagger$             & {\bf 0.08}            & {\bf 0.5}                & {\bf 0.8}             & OK & violated \\
\cite{werner2003social}     & $>2.25$                           & $\geq 3.34$           & {\bf 0.61}               & {\bf 1.37}            &    & \\
\cite{seer2014validating}   & {\bf 0.16 $\pm$ 0.01}           & {\bf 4.16 $\pm$ 0.65} & {\bf 0.5} $^\ddagger$    & $<0.087$              &    &\\
\cite{seer2014validating}   & {\bf 0.45 $\pm$ 0.41}             & {\bf 13.5 $\pm$ 18.4} & {\bf 0.5} $^\ddagger$    & $<0.43$               &    &\\
\cite{hoogendoorn2007}$^1$  & {\bf 12.0 $\pm$ 0.2}            & {\bf 0.16 $\pm$ 0.08} & {\bf 1.09 $\pm$ 0.35}    & {\bf 1.34 $\pm$ 0.21} & OK & violated \\
\hline
\end{tabular}
\caption{Bold face marks value from literature; normal face is computed with equations (\ref{eq:cond1}) and (\ref{eq:cond2}). Where one parameter was calculated from another one where a range is given the range was utilized such that the parameter space for the derived parameter is as large as possible. $^\dagger$: from {\bf 2000N}/75kg. $^\ddagger$: Assumption and input for calibration. $^1$: The simplified Nomad model discussed in said publication is identical to the circular specification of the Social Force Model, except that the radius of pedestrians is not stated explicitly in the exponent. If -- what is not stated explicitly in said contribution -- ``distance'' for the Nomad model means body surface to body surface the parameter meanings are identical, while if it means center to center distance the value of parameter $A$ would change by a factor $e^{-2R/B}$ and even equation (\ref{eq:cond1}) could be violated.
}
\label{tab:literaturecomparison}
\end{table}

\subsection{Elliptical specification II in comparison}
For elliptical specification II a rigid analysis is much more difficult than for the circular specification since the elliptical specification II does not only depend on the distance of the two pedestrians but also on their relative velocity. Still one can estimate the effect of the added consideration of relative velocity on the occurrence of oscillations. In the elliptical specification II of the Social Force Model \cite{Johansson2007} the force from pedestrian $\beta$ on pedestrian $\alpha$ is defined as
\begin{eqnarray}
\vec{f}_\alpha(\vec{r}_\alpha,\vec{r}_\beta,\dot{\vec{r}}_\alpha,\dot{\vec{r}}_\beta) &=&%
   w(\theta_{\alpha\beta}(\vec{d}_{\alpha\beta},\dot{\vec{r}}_\alpha)) \vec{g}(\vec{d}_{\alpha\beta},\dot{\vec{d}}_{\alpha\beta})\\
w(\theta_{\alpha\beta}(\vec{d}_{\alpha\beta},\dot{\vec{r}}_\alpha))                  &=&%
   \lambda_\alpha + (1-\lambda_\alpha) \frac{1+\cos(\theta_{\alpha\beta}(\vec{d}_{\alpha\beta},\dot{\vec{r}}_\alpha))}{2}\\
\cos(\theta_{\alpha\beta}(\vec{d}_{\alpha\beta},\dot{\vec{r}}_\alpha))               &=&%
   -\frac{\dot{\vec{r}}_\alpha \cdot \vec{d}_{\alpha\beta}}{|\dot{\vec{r}}_\alpha| |\vec{d}_{\alpha\beta}|}\\
\vec{d}_{\alpha\beta}                                                                   &=&%
   \vec{r}_\alpha-\vec{r}_\beta\\
\vec{g}(\vec{d}_{\alpha\beta},\dot{\vec{d}}_{\alpha\beta})                           &=&%
   -\vec{\nabla}_{\vec{d}_{\alpha\beta}}V_{\alpha\beta}(\vec{d}_{\alpha\beta},\dot{\vec{d}}_{\alpha\beta})\\
V_{\alpha\beta}(\vec{d}_{\alpha\beta},\dot{\vec{d}}_{\alpha\beta}) &=&%
   A_\alpha B_\alpha e^{-\frac{b_{\alpha\beta}(\vec{d}_{\alpha\beta},\dot{\vec{d}}_{\alpha\beta})}{B_\alpha}}\\
b_{\alpha\beta}(\vec{d}_{\alpha\beta},\dot{\vec{d}}_{\alpha\beta}) &=&%
   \frac{1}{2}\sqrt{(|\vec{d}_{\alpha\beta}|+|\vec{d}_{\alpha\beta}+\dot{\vec{d}}_{\alpha\beta}\Delta t_\alpha|)^2-(\dot{\vec{d}}_{\alpha\beta}\Delta t_\alpha)^2}
\end{eqnarray}
where $\vec{r}$ gives the position of pedestrians and $A$, $B$, $\lambda$, and $\Delta t$ are model parameters. Pedestrians' positions and derived properties are time dependent, other properties are constant. In one dimension and with pedestrians facing each other this simplifies to
\begin{eqnarray}
\cos(\theta_{\alpha\beta})                              &=& 1\\
w(\theta_{\alpha\beta})                                 &=& 1\\
d_{\alpha\beta}                                         &=& x_\alpha-x_\beta \text{ w.l.o.g. assumed to be $>0$}\\
f_\alpha(x_\alpha,x_\beta,\dot{x}_\alpha,\dot{x}_\beta) &=& g(d_{\alpha\beta},\dot{d}_{\alpha\beta}) = -\frac{d}{dd_{\alpha\beta}}V_{\alpha\beta}(d_{\alpha\beta},\dot{d}_{\alpha\beta})\\
V_{\alpha\beta}(d_{\alpha\beta},\dot{d}_{\alpha\beta})  &=& A_\alpha B_\alpha e^{-\frac{b_{\alpha\beta}(d_{\alpha\beta},\dot{d}_{\alpha\beta})}{B_\alpha}}\\
b_{\alpha\beta}(d_{\alpha\beta},\dot{d}_{\alpha\beta})  &=& \frac{1}{2}\sqrt{(d_{\alpha\beta}+|d_{\alpha\beta}+\dot{d}_{\alpha\beta}\Delta t_\alpha|)^2-(\dot{d}_{\alpha\beta}\Delta t_\alpha)^2}\\
                                                        &=& 0 \text{ for $(d_{\alpha\beta}+\dot{d}_{\alpha\beta}\Delta t_\alpha)\leq0$}\\
                                                        &=&\sqrt{d^2_{\alpha\beta}+d_{\alpha\beta}\dot{d}_{\alpha\beta}\Delta t_\alpha} \text{ otherwise}
\end{eqnarray}
Therefore the force is either zero if $(d_{\alpha\beta}+\dot{d}_{\alpha\beta}\Delta t_\alpha)\leq0$ or it is
\begin{eqnarray}
f_\alpha(x_\alpha,x_\beta,\dot{x}_\alpha,\dot{x}_\beta) &=& g(d_{\alpha\beta},\dot{d}_{\alpha\beta})\\
 &=& A_\alpha\frac{2d_{\alpha\beta}+\dot{d}_{\alpha\beta}\Delta t_\alpha}{2\sqrt{d^2_{\alpha\beta}+d_{\alpha\beta}\dot{d}_{\alpha\beta}\Delta t_\alpha}} e^{-\frac{\sqrt{d^2_{\alpha\beta}+d_{\alpha\beta}\dot{d}_{\alpha\beta}\Delta t_\alpha}}{B_\alpha}} \label{eq:20071d}\\
  &=& A_\alpha \frac{\bar{d}_a}{\bar{d}_g}e^{-\frac{\bar{d}_g}{B_\alpha}} \label{eq:20071d2}
\end{eqnarray}
with $\bar{d}_a$ being the arithmetic and $\bar{d}_g$ the geometric mean of current and projected distance ($d_{\alpha\beta}$ resp. $d_{\alpha\beta}+\dot{d}_{\alpha\beta}\Delta t_\alpha$). It can directly be seen that for large distances equation (\ref{eq:20071d}) reduces approximately to the circular specification from \cite{helbing2000simulating} which depends only on distance.

For a pedestrian $\alpha$ approaching another pedestrian $\beta$ from positive $x$ values $d$ is positive and $\dot{d}$ is negative. Therefore and because of the inequality of arithmetic and geometric means it holds $d_{\alpha\beta}+\dot{d}_{\alpha\beta}\Delta t_\alpha<\bar{d}_g<\bar{d}_a<d_{\alpha\beta}$. Then obviously as long as $d>-\dot{d}\Delta t$ the force as in equation (\ref{eq:20071d}) is larger compared to the circular specification and consequently pedestrian $\alpha$ will not overshoot for certain parameter choices where he does so with the circular specification.

In case pedestrian $\alpha$ overshoots over the equilibrium point and turns around it would be desirable that the force is not larger but smaller than with the circular specification. However, since in this case $\dot{d}$ becomes positive it holds that $d_{\alpha\beta}<\bar{d}_g<\bar{d}_a<d_{\alpha\beta}+\dot{d}_{\alpha\beta}\Delta t_\alpha$ and therefore in equation (\ref{eq:20071d2}) the exponential factor gives a smaller value than with the circular specification, yet the fraction factor has a value $>1$ and may for large values of parameter $B$ outweigh the damping effect from the modification in the exponential function. There are three indications that also in this phase of pedestrian $\alpha$'s movement and in general oscillations are suppressed with elliptical specification II: first, for large values of parameter $B$ already in the circular specification there are no oscillations as equation (\ref{eq:cond2}) tells us. This means that where in an isolated view on the ``way back'' the problem is most pronounced the system may actually not even evolve to the point that it exists.

Second, one can expand equation (\ref{eq:20071d}) in a series for small values of $\dot{d}_{\alpha\beta}\Delta t_\alpha/d_{\alpha\beta}$:
\begin{equation}
f_\alpha(x_\alpha,x_\beta,\dot{x}_\alpha,\dot{x}_\beta) \approx A_\alpha e^{-\frac{d_{\alpha\beta}}{B_\alpha}}\left(1-\frac{d_{\alpha\beta}}{B_\alpha}\frac{\dot{d}_{\alpha\beta}\Delta t_\alpha}{d_{\alpha\beta}}\right)\label{eq:series20071d}
\end{equation}
In the moment pedestrian $\alpha$ turns around it is $\dot{d}_{\alpha\beta}=0$ and therefore circular and elliptical specification II yield an identical force. Starting to move backward with now positive $\dot{d}_{\alpha\beta}$ equation (\ref{eq:series20071d}) tells us that elliptical specification II yields smaller forces with the difference to circular specification leveled for $B_\alpha \rightarrow \infty$.

Third, with the -- admittedly arguable -- additional assumptions that
\begin{eqnarray}
\dot{d}_{\alpha\beta}\Delta t_\alpha &<<& d_{\alpha\beta}\\
\dot{d}_{\alpha\beta}\Delta t_\alpha &<<& B_\alpha
\end{eqnarray}
one cannot only expand for small $\xi(t)$, but also reduce the complexity of equation (\ref{eq:20071d}) with regard to $\dot{d}_{\alpha\beta}$. Various approximate forms of that equation can be derived in this way of which one is analytically solvable {\em and} contains parameter $\Delta t_\alpha$ (omitting the indices $\alpha$):
\begin{equation}
\ddot{\xi}(t) + \frac{1}{\tau}\left(1+\frac{v_0\Delta t}{B}\right)\dot{\xi}(t)+\frac{v_0}{B\tau}\xi(t)=0
\end{equation}
where in comparison to equation (\ref{eq:dho}) just the factor before $\dot{\xi}(t)$ has changed. This leads to a less strict requirement for avoiding oscillations:
\begin{equation}
4v_0\tau\leq B\left(1+\frac{v_0\Delta t}{B}\right)^2
\end{equation}
than equation (\ref{eq:cond2}).

Finally a note on the case $d+\dot{d}\Delta t < 0$ where the force is zero: if pedestrian $\alpha$ starts to approach $\beta$ from a sufficiently large distance and with a typical pedestrian walking speed it will be $d+\dot{d}\Delta t > 0$. Since equation (\ref{eq:20071d}) diverges to positive values at $d+\dot{d}\Delta t\rightarrow 0^{+}$ the pedestrian will slow down and eventually be at rest or turn around before $d<-\dot{d}\Delta t$. For a departing pedestrian $\alpha$ it is $\dot{d}>0$ and thus always $d+\dot{d}\Delta t > 0$. Only when a simulation is initiated with $d+\dot{d}\Delta t<0$ it may yield dynamics that do not fit into this line of argumentation; compare \cite{schadschneider1998garden}. Such extrinsically prepared ``Garden of Eden'' states can be dealt with for example by a model extension (if one does not simply want to exclude them by construction).

\subsection{Simulations}
For realistic applications it makes sense to choose the parameters such that no oscillations occur. However, to verify a computer implementation of the Social Force Model it can be interesting to use parameters just around the critically damped conditions and check if oscillations do and do not occur according to expectations. Model specific validation work can amend validation tests like those in the RiMEA test cases \cite{Rimea2009} which usually are formulated model independently. In this subsection such a model specific verification process will be carried out exemplarily utilizing PTV Viswalk \cite{Viswalk6}. 

\begin{figure}[htbp]%
\center
\includegraphics[width=0.612\columnwidth]{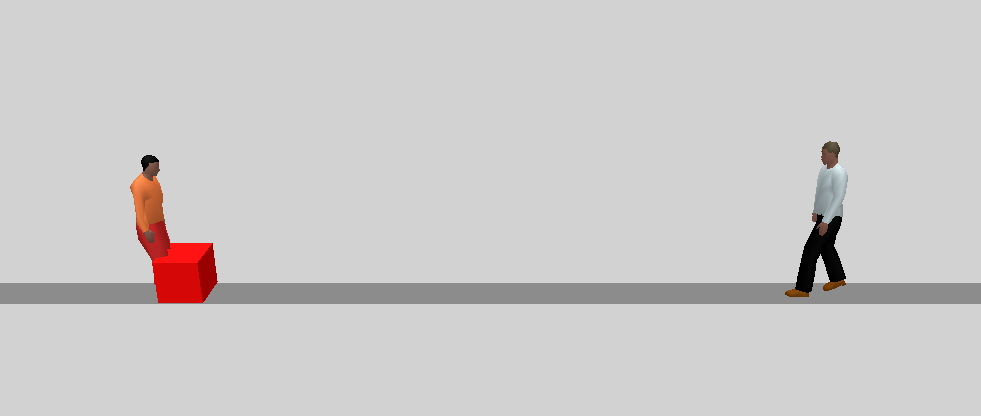}%
\caption{Simulation scenario.}%
\label{fig:simscen}%
\end{figure}

Figure \ref{fig:simscen} shows how the one-dimensional setting has been implemented in the two-dimensional simulation model. The walking area is modeled to be 0.55 m wide. A pedestrian facing into positive x direction is stopped by a red traffic signal at x=0. Then a second pedestrian is set into the model at x=52 m moving into negative x direction. 

\begin{table}[htbp]
\center
\begin{tabular}{c|ccc}
$\tau$ & distance & expected & difference\\ \hline
0.7 & 0.4556 & 0.4570 & 0.0014 \\
0.8 & 0.4821 & 0.4837 & 0.0016 \\
0.9 & 0.5044 & 0.5072 & 0.0028 \\
1.0 & 0.5269 & 0.5283 & 0.0014 \\
1.2 & 0.5645 & 0.5648 & 0.0002 \\
1.5 & 0.6073 & 0.6094 & 0.0021 \\
2.0 & 0.6655 & 0.6669 & 0.0014 \\
3.0 & 0.7483 & 0.7480 &-0.0003 \\
4.0 & 0.8044 & 0.8056 & 0.0011 \\
5.0 & 0.8482 & 0.8502 & 0.0020 \\\hline
\end{tabular}
\caption{Stand still distance [m] in dependence of parameter $\tau$ [s], with $A=1.6$ $m/s^2$, $B=0.2$ $m$ and $v_0=1.5$ $m/s$. Column ``expected'' shows the values according to equation (\ref{eq:ds}). }
\label{tab:results2}
\end{table}

Parameter settings: since $A\_soc\_mean$ controls the contribution of elliptical II specification it has been set to zero (in the remainder ''$A$´´ refers to ''$A\_soc\_iso$´´ in the software; for the pedestrian on the left it has been set to zero as well); $\lambda=1.0$ to minimize effects from the in fact two-dimensional nature of the simulation; stochastic noise has been set to zero . The value of the noise parameter has been set to zero for both pedestrians to not unnecessarily blur the results.

\begin{table}[htbp]
\center
\begin{tabular}{c|ccc||c|ccc}
$B$ & distance & expected & difference & $B$ & dist. & exp. & diff.\\ \hline
 0.1 &  0.5831 &  0.5847 &  0.0016 &  4.0 &  3.2870 &  3.2880 &  0.0010 \\
 0.2 &  0.6524 &  0.6540 &  0.0016 &  6.0 &  4.6734 &  4.6743 &  0.0008 \\
 0.3 &  0.7240 &  0.7233 & -0.0007 &  9.0 &  6.7530 &  6.7537 &  0.0007 \\
 0.5 &  0.8590 &  0.8620 &  0.0029 & 12.0 &  8.8326 &  8.8332 &  0.0005 \\
 1.0 &  1.2073 &  1.2085 &  0.0013 & 18.0 & 12.9917 & 12.9920 &  0.0003 \\
 2.0 &  1.9005 &  1.9017 &  0.0012 & 24.0 & 17.1509 & 17.1509 &  0.0000 \\ \hline
\end{tabular}
\caption{Stand still distance [m] in dependence of parameter $B$ [m], with $A=2.0$ $m/s^2$, $\tau=1.5$ $s$ and $v_0=1.5$ $m/s$. Column ``expected'' shows the values according to equation (\ref{eq:ds}). The difference between theoretical expectation and simulation is in all cases below 3 mm.}
\label{tab:results3}
\end{table}

\begin{figure}[htbp]%
\center
\includegraphics[width=0.45\columnwidth]{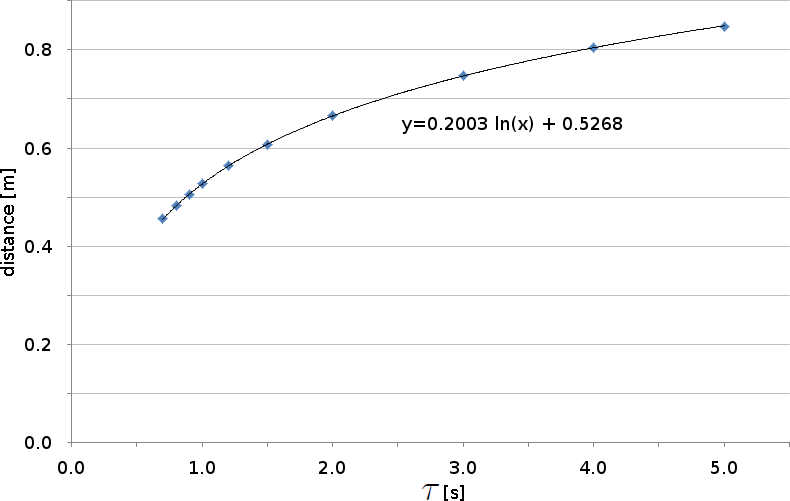}\hspace{12pt}%
\includegraphics[width=0.45\columnwidth]{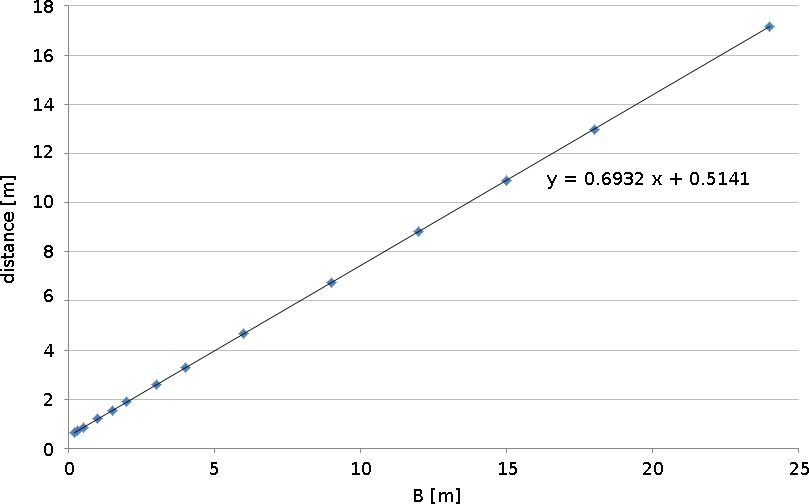}%
\caption{Left: Visualization of the data of table \ref{tab:results2}. The expectation for the regression curve would be $y=0.2000\ln(x)+0.5283$. Right: Visualization of the data of table \ref{tab:results3}. The expectation for the regression curve would be $y = 0.6931 x + 0.5154$.}%
\label{fig:results2}%
\end{figure}

At first we investigate where the second pedestrian comes to rest. According to equation (\ref{eq:ds}) this depends on the values of the parameters $B\_soc\_iso$, $A\_soc\_iso$, $v_0$, $\tau$, and $R$. The latter was set to be 0.2577 $m$. Keeping $A\_soc\_iso$ and $v_0$ constant and increasing the value of $\tau$ in steps of $0.1$ s the second pedestrian does not pass through the first one for the first time at a value $\tau=0.6$ s and both do not overlap visually for $\tau \geq 0.8$ s. At $\tau=0.9375$ s when $A \tau = v_0$ the distance of the central points of both pedestrians is 0.5135 m which comes 2 mm close to $2R$. Table \ref{tab:results2} and figure \ref{fig:results2} show values for the stand still distance in dependence of some values for parameter $\tau$ with all other parameters kept constant. The theoretical expectation is met well in all cases. 

Table \ref{tab:results3} and figure \ref{fig:results2} show values for the stand still distance in dependence of some values for parameter $B\_soc\_iso$ with all other parameters kept constant. The theoretical expectation is met well in all cases. 

\begin{figure}[htbp]%
\center
\includegraphics[width=0.45\columnwidth]{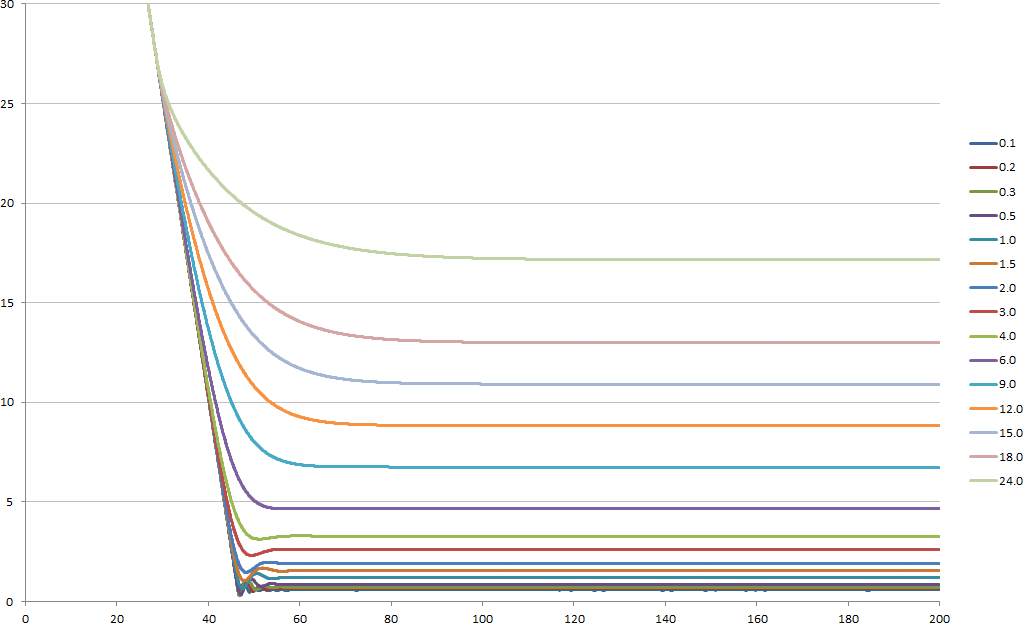} \hspace{12pt}%
\includegraphics[width=0.45\columnwidth]{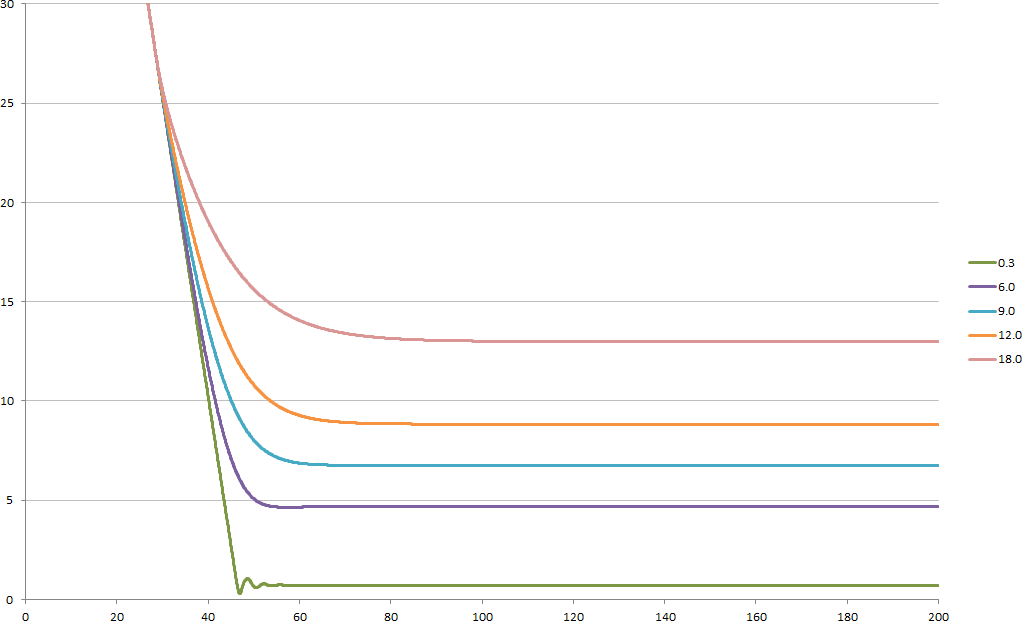}%
\caption{Left: Position of approaching pedestrian over time for various values of parameter $B$. Right: As $A=2.0$ $m/s^2$, $v_0=1.5$ $m/s$, $\tau=1.5$ $s$ the system is critically damped for $B=9.0$ $m$. For increasing $B$ the oscillations get smaller and vanish for $B=9.0$.}%
\label{fig:results4}%
\end{figure}

\begin{figure}[htbp]%
\center
\includegraphics[width=0.45\columnwidth]{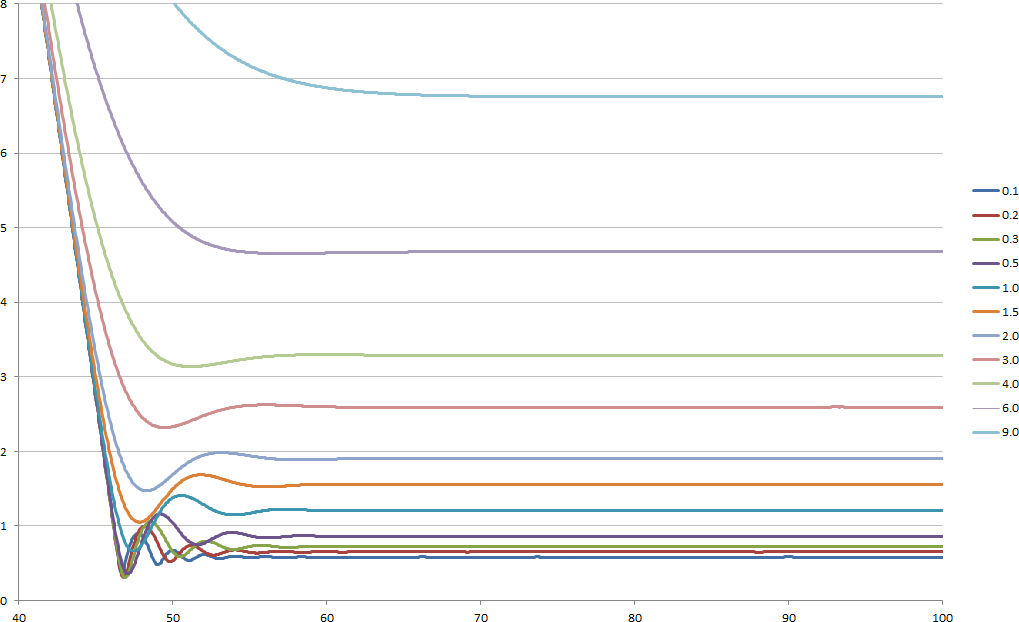} \hspace{12pt}%
\includegraphics[width=0.45\columnwidth]{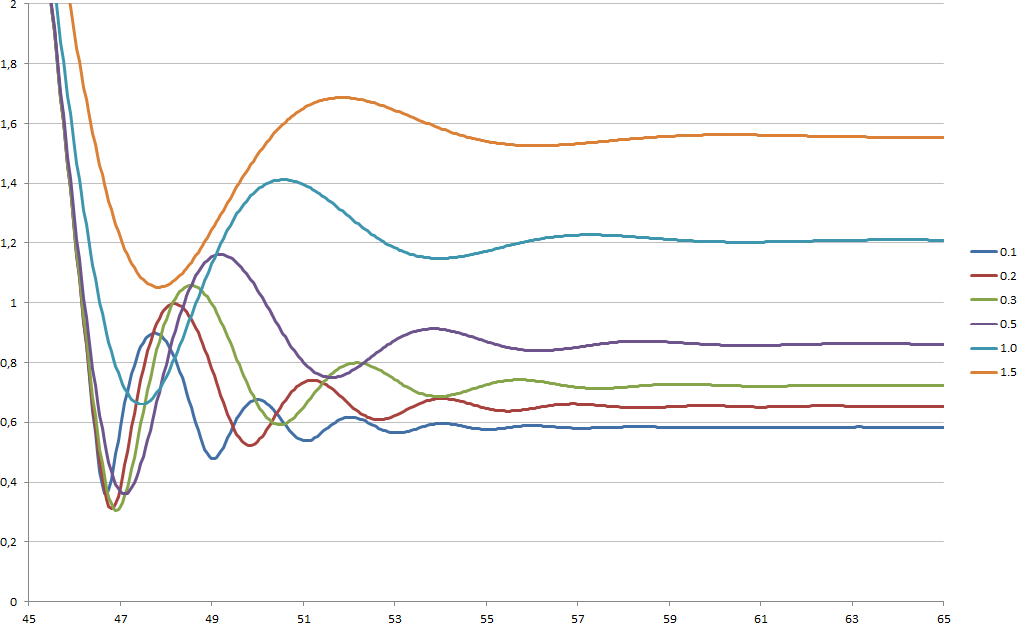}%
\caption{Left: Position of approaching pedestrian vs. time for under damped and critical cases with regard to the value of $B$. Right: Zoom to the region of oscillations.}%
\label{fig:results7}%
\end{figure}

The stand still distances of table \ref{tab:results3} are unrealistically high for all but the smallest values for parameter $B$. We have chosen the parameters in this way not to scan for realistic parameter values, but because with these parameters one can well demonstrate that for certain cases (small values for $B$) there are oscillations and in others (large values for $B$) there are none. Figures \ref{fig:results4} and \ref{fig:results7} show the time evolution of the position of the approaching pedestrian for various values of parameter $B$.

Neglecting for the under damped case the damping exponential function the approximately expectated time distance $T_r$ between two reversal points in the under damped cases is:
\begin{equation}
T_r=\frac{\pi}{\sqrt{\frac{v_0}{B\tau}-\frac{1}{4\tau^2}}}
\end{equation}
Table \ref{tab:freq} shows a comparison between actual and expected $T_r$ for various values for $B$ which generate under damped behavior.

\subsection{Two pedestrians approaching each other}

If two pedestrians approach each other in one dimension they may come to a stand still at an equilibrium distance or they might jointly (and with an equilibrium distance) move into one of the two possible directions. If all parameters ($v_0$, $A$, $B$, and $\tau$) are identical for both pedestrians one can rewrite the coupled equations of motion of both pedestrians as one equation for the movement of the center of mass and one equation for the relative motion. Latter one can again be approximated by an oscillator equation. Compared to the case above one has an additional factor 2 at the friction term implying that the system is critically or over damped for
\begin{equation}
8 v_0\tau \leq B
\end{equation}
which leads to the conclusion that the case of two mutually approaching pedestrians sets a stronger restriction on parameter choice than the case when one pedestrian is approaching another who is standing, at least if oscillations are to be avoided. The 2007 variant of the Social Force Model in this case suppresses oscillations even more than when one of the two pedestrians is standing still.

\begin{table}[htbp]
\center
\begin{tabular}{c|ccc}
$B$ & number of reverses & simulation & approximate expectation\\ \hline
 0.1 &  11               &    1.035   & 0.999 \\
 0.2 &   7               &    1.442   & 1.421 \\
 0.3 &   3               &    1.800   & 1.750 \\
 0.5 &   3               &    2.250   & 2.286 \\
 1.0 &   2               &    3.100   & 3.332 \\
 1.5 &   2               &    4.000   & 4.215 \\
 2.0 &   2               &    4.850   & 5.038 \\
 3.0 &   2               &    6.550   & 6.664 \\ \hline
\end{tabular}
\caption{Comparison of simulated and approximately expected time [$s$] between reversals.}
\label{tab:freq}
\end{table}

\section{Conclusions}
It could be shown that the Social Force Model as proposed in 1995 (elliptical specification I) and 2000 (circular specification) in a special and one-dimensional case and around equilibrium reduces to the equations of the damped harmonic oscillator. This implies that indeed one has to be careful not to choose parameters with which the pedestrians' trajectories yield unrealistic results. However, at the same time it means that there are parts in parameter space in which there are not only just small oscillations, but where there are exactly no oscillations. A look at parameter values found in literature for the circular specification shows that parameters deemed to yield realistic results in certain calibration scenarios do or may produce oscillating behavior unless the desired walking speed(s) are set to rather small values.

The equations of the Social Force Model as of 2007 (elliptical specification II) are not as easily treated analytically. Still in a discussion of its equations it could be argued that -- compared to the circular specification -- oscillations are clearly suppressed. Elliptical specification II from 2007 therefore appears to be superior to the two preceding variants also for the reasons discussed in this paper (this was found already in the paper from 2007 but for different reasons). It is therefore a good idea to either simply use elliptical specification II or combine it with one of the earlier variants (e.g. by simply adding the forces) to reduce the risk of oscillations.

The phenomenon of oscillations was used to verify a specific computer implementation of the Social Force Model. It was possible to show that it reproduces the expected results. This comparison can be seen as an attempt to falsify either of the two -- theoretical expectations and software implementation -- and the attempt failed, no potential issues could be found. The method can be applied to verify any implementation of the Social Force Model variants of 1995 or 2000 and it is generally an example of model specific verification. The tests carried out in this contribution are just examples. The phenomenon of oscillations bears the potential to formulate further tests.

\section{Acknowledgments}
I thank Mohcine Chraibi and Normen Rochau for useful discussions.


\section{References}

\bibliographystyle{utphys2011}
\bibliography{oscillations}

\begin{appendix}

\section{Proof that a pedestrian heading for a destination point never comes to rest} \label{app:convergence}
In this section we show that
\begin{equation}
t_n=\sum_{i=1}^{n}\Delta t
\end{equation}
diverges for $n\rightarrow \infty$.

\begin{figure}[htbp]%
\center
\includegraphics[width=0.5\columnwidth]{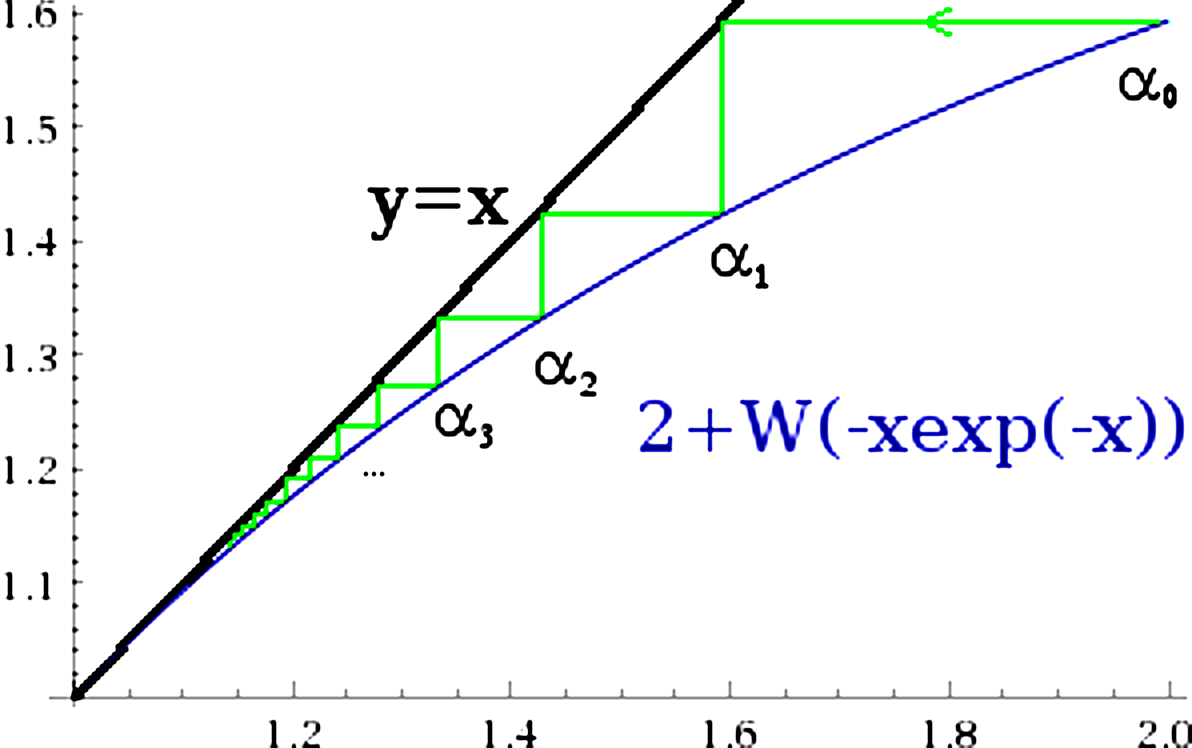}%
\caption{Plot of $2+W_0\left(-x e^{-x}\right)$ for $1\leq\alpha\leq 2$ and a visualization of the evolution of the $\alpha_n$; basically created with \cite{wolframalpha2}.}%
\label{fig:W-xe-x}%
\end{figure}

First step: We proof that $\alpha_{n+1}<\alpha_n$:
we begin with $\alpha_0=2$ and with equation (\ref{eq:alphanp1})
\begin{equation}
\alpha_{n+1}=2+W_0\left(-\alpha_n e^{-\alpha_n}\right)
\end{equation}
For all $1<z\leq2$ it holds that
\begin{eqnarray}
-ze^{-z}&<&0\\
\frac{\partial}{\partial z}-ze^{-z} &>& 0
\end{eqnarray}
and for all $-1/e<z<0$
\begin{eqnarray}
W_0(z)&<&0\\
\frac{\partial}{\partial z} W_0(z) &>& 0
\end{eqnarray}
Therefore for all $1<z<2$ it holds that
\begin{eqnarray}
W_0(-ze^{-z})&<&0\\
\frac{\partial}{\partial z}W_0(-ze^{-z}) > 0
\end{eqnarray}
From that and with $\alpha_0=2$ it follows that
\begin{equation}
\alpha_1=2+W_0(-\alpha_0e^{-\alpha_0})<\alpha_0
\end{equation}
From equation (\ref{eq:alphanp1}) it follows that
\begin{equation}
\alpha_n-\alpha_{n+1}=W_0(-\alpha_{n-1}e^{-\alpha_{n-1}})-W_0(-\alpha_ne^{-\alpha_n})
\end{equation}
and from it that if $\alpha_{n-1}>\alpha_{n}$ also $\alpha_{n}>\alpha_{n+1}$. See figure \ref{fig:W-xe-x} for a plot of $2+W_0\left(-z e^{-z}\right)$ and the visualization of the evolution of the $\alpha_n$ which allows to see this first step very easily.

Second step: We verify that $\Delta t_{n} > 0$ (physically this is obvious, mathematically it needs to be shown; this step in effect is a check on potential errors done before):
Since
\begin{equation}
\alpha_n=-W_1\left(-\alpha_n e^{-\alpha_n}\right)
\end{equation}
and because for $-1/e<z<0$ it holds that
\begin{equation}
W_0(z)>W_1(z)
\end{equation}
it results for the $\Delta t_n$ according to equation (\ref{eq:Dt})
\begin{equation}
\frac{\Delta t_{n+1}}{\tau} = \alpha_n + W_0\left(-\alpha_n e^{-\alpha_n}\right) = W_0\left(-\alpha_n e^{-\alpha_n}\right) - W_1\left(-\alpha_n e^{-\alpha_n}\right) > 0
\end{equation}

Third step: we proof that $\Delta t_{n+1} < \Delta t_n$. Equations (\ref{eq:Dt}) and (\ref{eq:alphanp1}) read
\begin{eqnarray}
\frac{\Delta t_{n+1}}{\tau} &=& \alpha_n + W_0(-\alpha_ne^{-\alpha_n}) \ \\
\alpha_n &=& 2 + W_0(-\alpha_{n-1}e^{-\alpha_{n-1}}). \label{eq:alphanalphannm1}
\end{eqnarray}
From them it follows that
\begin{equation}
\frac{\Delta t_{n+1}}{\tau} = \alpha_{n+1} + \alpha_n - 2 \label{eq:alphaalpha}
\end{equation}
thus the $\Delta t_{n}$ show the same behavior as the $\alpha_n$ and with $\alpha_{n+1} < \alpha_n$ also $\Delta t_{n+1} < \Delta t_n$.

Fourth step: if $\Delta t_{n+1}/\Delta t_n$ for large $n$ would approach a value smaller than 1 the sum would converge, only if it approaches 1 it may diverge. From (\ref{eq:alphaalpha}) it follows that

\begin{equation}
\frac{\Delta t_{n+1}}{\Delta t_n} =  1 - \frac{\alpha_{n-1}-\alpha_{n+1}}{\alpha_{n-1}+\alpha_n-2}\label{eq:tbyt}
\end{equation}

To compute the limit for this for large $n$ we have to expand $W_0(-\alpha_n e^{-\alpha_n})$ into a series.

For convenience we write
\begin{equation}
\beta_n=\alpha_n-1
\end{equation}
knowing that this is $\beta_n=u_n/v_0$ and that these $\beta_n$ will approach 0 with increasing $n$.

\begin{figure}[htbp]%
\center
\includegraphics[width=0.5\columnwidth]{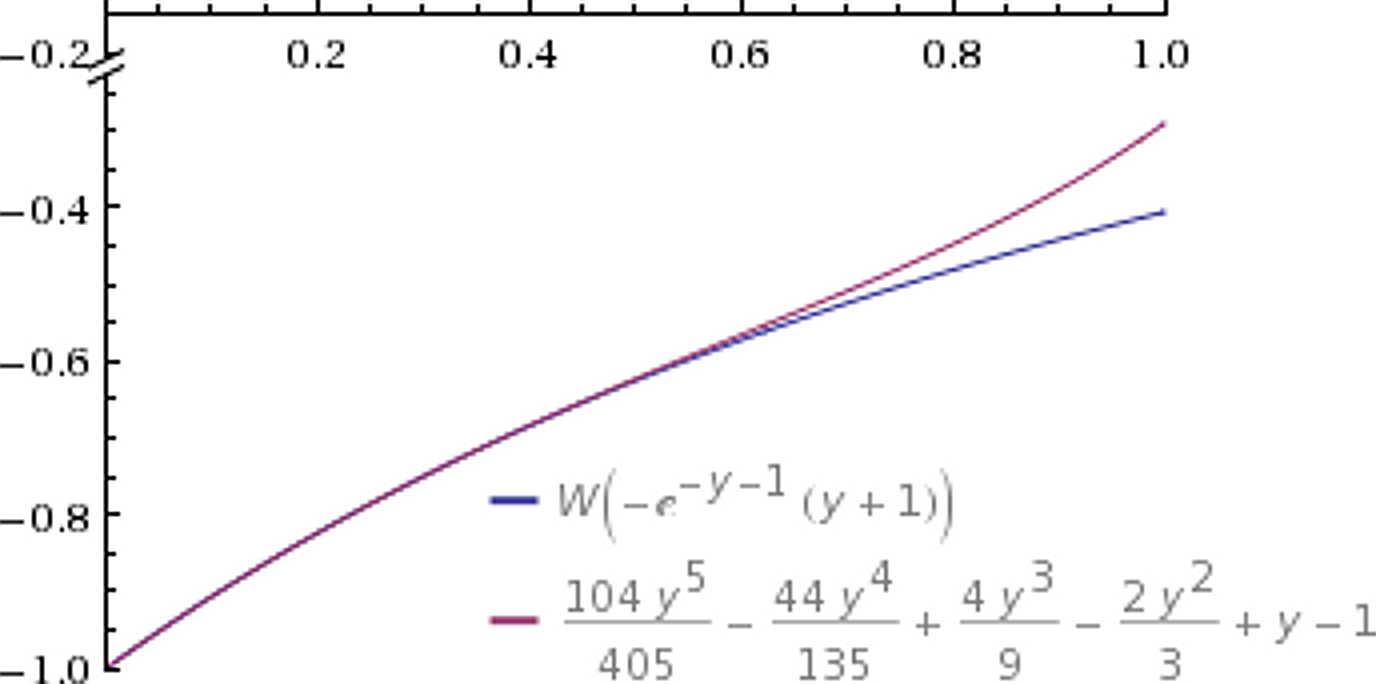}%
\caption{Plot comparing $W(-(1+y)$ $\exp(-(1+y)))$ and its approximation \cite{wolframalpha3}.}%
\label{fig:approxW}%
\end{figure}

By computing the right side limit towards 0 of 
\begin{equation}
W_0(-\alpha_{n}e^{-\alpha_{n}})=W_0(-(1+\beta_n)e^{-(1+\beta_n)})
\end{equation}
and its derivatives one can write
\begin{equation}
W_0(-(1+\beta_n)e^{-(1+\beta_n)})= -1 + \beta_n -\frac{2}{3}\beta_n^2 + \frac{4}{9}\beta_n^3 - \frac{44}{135}\beta_n^4 + \frac{104}{405}\beta_n^5 +O(\beta_n^6) \label{eq:Wtaylor}
\end{equation}
compare figure \ref{fig:approxW}.

Considering only terms to second order
\begin{eqnarray}
\beta_n&\approx& \beta_{n-1} - \frac{2}{3} \beta_{n-1}^2\\
\beta_{n+1}&\approx& \beta_{n-1} - \frac{4}{3} \beta_{n-1}^2
\end{eqnarray}

Using these approximations in equation (\ref{eq:tbyt}) gives for large $n$
\begin{equation}
\frac{\Delta t_{n+1}}{\Delta t_n} \approx 1 - \frac{2 \beta_{n-1}}{3 - \beta_{n-1}}
\end{equation}
which approaches 1 with $n\rightarrow \infty$ where $\beta_{n-1}\rightarrow 0$.

Thus, like with the harmonic series $h_m$ the ratio of subsequent terms approaches 1. 
We note that equation (\ref{eq:alphaalpha}) implies $\frac{\Delta t_{n+1}}{\tau} = \beta_{n+1} + \beta_n$ and therefore
\begin{equation}
\frac{t_{n\rightarrow \infty}}{\tau} = \alpha_0 - 1 + 2\sum_{n=1}^\infty \beta_n
\end{equation}
the sum of the $\Delta t_n$ depends trivially on the sum of the $\beta_n$ and especially their convergence behavior is the same. In the last step we prove that if $\beta_n>h_m$ then also $\beta_{n+1}>h_{m+1}$ and therefore beyond some $n,m$ the harmonic series is a lower estimate of the $\beta_n$ implying that with the harmonic series also the sum of $\beta_n$ diverges.

Assume that for some $n$, $\beta_n$ can be written as
\begin{equation}
\beta_n = \frac{q}{m}
\end{equation}
with $m \in \mathbb{N}$, $m>>1$ and $1\leq q < m/(m-1)$, i.e. $\beta_n$ is placed between two elements of the harmonic series. This requires just that there is some $0<\beta_n<1$. This is obviously the case.

So, can there be a $\beta_{n+1}<1/(m+1)$?
\begin{equation}
\beta_{n+1}-\frac{1}{m+1} = \frac{q}{m} - \frac{2}{3}\frac{q^2}{m^2} - \frac{1}{m+1}
\end{equation}
A lower estimate of the right side is if at the positive term $q$ is replaced by its minimum value $q=1$ and at the negative term by its upper limit $q=m/m-1$:
\begin{eqnarray}
\beta_{n+1}-\frac{1}{m+1} &>& \frac{1}{m} - \frac{2}{3}\frac{1}{(m-1)^2} - \frac{1}{m+1}\\
&>&\frac{m^2-8m+3}{3m(m-1)^2(m+1)}\\
&>&0 \forall (m\geq 8)
\end{eqnarray}
Thus, if there is a $\beta_{n_0}<1/8$ -- it is, as can be seen in table \ref{tab:results1}) -- it will hold for all $\beta_{n>n_0}$ that if $\beta_n>h_m$ then also $\beta_{n+1}>h_{m+1}$. Therefore the harmonic series is a lower estimate for the series of $\beta_n$. Thus the series of the $\beta_n$ diverges and with it the series of the $\Delta t_n$.

\end{appendix}

\end{document}